\documentclass[twocolumn,showpacs,preprintnumbers,amsmath,amssymb,superscriptaddress]{revtex4}
\usepackage{graphicx,bm}
\def\be{\begin{equation}}
\def\ee{\end{equation}}
\def\bea{\begin{eqnarray}}
\def\eea{\end{eqnarray}}
\def\nnb{\nonumber}
\def\bbuildrel#1_#2^#3{\mathrel{\mathop{\kern 0pt#1}\limits_{#2}^{#3}}}
\newcommand{\gsim}{\;\rlap{\lower 3 pt \hbox{$\mathchar \sim$}} \raise 2pt \hbox {$>$}\;}
\newcommand{\lsim}{\;\rlap{\lower 3 pt \hbox{$\mathchar \sim$}} \raise 2pt \hbox {$<$}\;}
\def\theequation{\arabic{section}.\arabic{equation}}
\newcommand{\newsection}[1]{\section{#1}\setcounter{equation}{0}}
\newcommand{\newappendix}[1]{\section*{#1}\setcounter{equation}{0}}
\newcommand{\scs}{\scriptscriptstyle}
\newcommand{\f}{\frac}

\begin{document}

\preprint{IFT-4/2012, TTP12-033, SI-HEP-2011-09}
\title{Tree-level contributions to {\boldmath $\bar B \to X_s \gamma$}}
\author{Maciej Kami\'nski}
\affiliation{Faculty of Physics, University of Warsaw, PL-00-681 Warsaw, Poland}
\author{Miko{\l}aj Misiak} 
\affiliation{Faculty of Physics, University of Warsaw, PL-00-681 Warsaw, Poland}
\affiliation{Institut f\"ur Theoretische Teilchenphysik, Karlsruhe Institute of Technology,
          D-76128 Karlsruhe, Germany}
\author{Micha{\l} Poradzi\'nski}
\affiliation{Faculty of Physics, University of Warsaw, PL-00-681 Warsaw, Poland}
\affiliation{Theoretische Physik 1, Naturwissenschaftlich-Technische Fakult\"at,
Universit\"at Siegen, D-57068 Siegen, Germany}
\begin{abstract}
Weak radiative decay $\bar B \to X_s \gamma$ is known to be a loop-generated
process. However, it does receive tree-level contributions from CKM-suppressed
$b \to u\bar u s\gamma$ transitions. In the present paper, we evaluate such
contributions together with similar ones from the QCD penguin operators. For a
low value of the photon energy cutoff $E_0 \simeq m_b/20$ that has often been
used in the literature, they can enhance the inclusive branching ratio by more
than 10\%. For $E_0 = 1.6\;$GeV or higher, the effect does not exceed $0.4\%$,
which is due to phase-space suppression. Our perturbative results contain
collinear logarithms that depend on the light quark masses $m_q$~
($q=u,d,s$). We have allowed $m_b/m_q$ to vary from 10 to 50, which
corresponds to values of $m_q$ that are typical for the constituent quark
masses. Such a rough method of estimation may be improved in the future with
the help of fragmentation functions once the considered effects begin to
matter in the overall error budget for ${\cal B}(\bar B \to X_s \gamma)$.
\end{abstract}
\maketitle

\newsection{Introduction \label{sec:intro}}

Weak radiative decay of the $B$ meson is an invaluable and well-established
means for constraining physics beyond the Standard Model (SM). Its
branching ratio has been measured to a few percent accuracy at the
$B$-factories~\cite{Chen:2001fj}. Theoretical calculations have acquired
similar precision~\cite{Misiak:2006zs}. The decay is generated dominantly by
the $b \to s \gamma$ transition that arises at one loop in the SM and its most
popular extensions. However, it receives also tree-level contributions from
the CKM-suppressed $b \to u\bar u s\gamma$ transitions. In the present paper,
we evaluate such contributions together with similar ones that originate from
the QCD penguin operators.

It is well known that the inclusive $\bar B \to X_s \gamma$ decay rate can be
approximated by its perturbative counterpart
\be \label{approx}
\Gamma(\bar B \to X_s \gamma)_{E_\gamma > E_0} \simeq
\Gamma(b \to X^p_s \gamma)_{E_\gamma > E_0}, 
\ee 
where $X^p_s$ stands for $s,sg,sq\bar{q},\ldots$ partonic states (with
$q=u,d,s$ only, as no charmed hadrons appear in $X_s$ by
definition). Deviations from Eq.~(\ref{approx}) appear as corrections when
$E_0$ is large ($E_0 \sim \f{1}{2} m_b$) but not too close to the endpoint ($m_b -
2E_0 \gg \Lambda \sim \Lambda_{\scs\rm QCD}$). For $E_0 = 1.6\,$GeV, the
corresponding non-perturbative uncertainty amounts to around $\pm
5\%$~\cite{Benzke:2010js}, while the known ${\cal O}(\Lambda^2/m_{b,c}^2)$
corrections~\cite{Bigi:1992ne,Buchalla:1997ky} are smaller than that.

The considered process is most conveniently analyzed in the framework of
an effective low-energy theory which arises after integrating out the
electroweak bosons and the top quark. The relevant effective weak
interaction Lagrangian reads
\be 
{\cal L}_{\rm weak} = 
 \f{4 G_F}{\sqrt{2}} \left[ V^*_{ts} V_{tb} \sum_{i=1}^{8} C_i P_i
                            +  V^*_{us} V_{ub} \sum_{i=1}^{2} C_i (P_i - P^u_i) \right],
\label{Leff2}
\ee                        
where $P_i$ denote either dipole-type or four-quark operators,
and $C_i$ stand for their Wilson coefficients.  
The operators are given by
\bea 
\begin{array}{rl}
P^u_1 ~= & (\bar{s}_L \gamma_{\mu} T^a u_L) (\bar{u}_L \gamma^{\mu} T^a b_L),\\[2mm]
P^u_2 ~= & (\bar{s}_L \gamma_{\mu}     u_L) (\bar{u}_L \gamma^{\mu}     b_L),\\[2mm]
P_1   ~= & (\bar{s}_L \gamma_{\mu} T^a c_L) (\bar{c}_L \gamma^{\mu} T^a b_L),\\[2mm]
P_2   ~= & (\bar{s}_L \gamma_{\mu}     c_L) (\bar{c}_L \gamma^{\mu}     b_L),\\[2mm]
P_3   ~= & (\bar{s}_L \gamma_{\mu}     b_L) \sum_q (\bar{q}\gamma^{\mu}     q),\\[2mm]
P_4   ~= & (\bar{s}_L \gamma_{\mu} T^a b_L) \sum_q (\bar{q}\gamma^{\mu} T^a q),\\[2mm]
P_5   ~= & (\bar{s}_L \gamma_{\mu_1}
                      \gamma_{\mu_2}
                      \gamma_{\mu_3}    b_L)\sum_q (\bar{q} \gamma^{\mu_1} 
                                                            \gamma^{\mu_2}
                                                            \gamma^{\mu_3}     q),\\[2mm]
P_6   ~= & (\bar{s}_L \gamma_{\mu_1}
                      \gamma_{\mu_2}
                      \gamma_{\mu_3} T^a b_L)\sum_q (\bar{q} \gamma^{\mu_1} 
                                                             \gamma^{\mu_2}
                                                             \gamma^{\mu_3} T^a q),\\[2mm]
P_7   ~= &  \f{e}{16\pi^2} m_b (\bar{s}_L \sigma^{\mu \nu}     b_R) F_{\mu \nu},\\[2mm]
P_8   ~= &  \f{g}{16\pi^2} m_b (\bar{s}_L \sigma^{\mu \nu} T^a b_R) G_{\mu \nu}^a.
\end{array}\nnb \label{physical}\\[-6mm]
\eea
Sums over $q$ in $P_{3,\ldots,6}$ include  all 
the active flavors $q=u,d,s,c,b$.
\begin{figure}[t]
\begin{center}
\includegraphics[width=7cm,angle=0]{fig1.ps}
\caption{\sf Tree-level Feynman diagrams for $b\rightarrow
s\bar{q}q\gamma$ at the LO. Black squares denote four-quark operator
insertions.\label{fig:tree}}
\end{center}
\vspace*{-3mm}
\end{figure}

Our goal in the present work is to evaluate ${\cal O}\left(\alpha_s^0\right)$ 
leading order (LO) contributions to the r.h.s. of Eq.~(\ref{approx}) that
originate from $b \to q\bar{q}s\gamma$ with $q=u,d,s$. They can be generated
either by the current-current operators $P^u_{1,2}$ or by
the QCD penguin ones $P_{3,\ldots,6}$. The corresponding
Feynman diagrams are shown in Fig.~\ref{fig:tree}.

At first glance, it may seem surprising that the considered LO effects have
not been evaluated so far while the analysis of other contributions has
already reached the ${\cal O}\left(\alpha_s^2\right)$ next-to-next-to-leading
order (NNLO) level~\cite{Misiak:2006zs}. Let us recall that the Wilson
coefficients of all the operators in Eq.~(\ref{physical}) acquire non-zero
values already at the LO once the QCD logarithms have been resummed using
renormalization group evolution from the electroweak scale $\mu_0 \sim M_W,
m_t$ down to the low-energy scale $\mu_b \sim m_b/2$. However, despite being
non-vanishing at the LO, Wilson coefficients of the QCD penguin operators
remain rather small $|C_{3,\ldots,6}(\mu_b)/C_7(\mu_b)|^2 <
|C_4(\mu_b)/C_7(\mu_b)|^2 \sim 0.1$, while the tiny CKM matrix element ratio
$\left|V_{us}^*V_{ub}/V_{ts}^*V_{tb}\right| \simeq 0.02$ makes the
current-current operators $P^u_{1,2}$ even more suppressed.  Moreover, when
the lower photon energy cutoff $E_0$ is at $1.6\,$GeV~\cite{Misiak:2006zs} or
higher~\cite{Chen:2001fj}, the considered tree-level contributions to the
branching ratio undergo severe phase-space suppression, which justifies
neglecting them at the leading and next-to-leading orders in $\alpha_s$ (see
App.~E of Ref.~\cite{Gambino:2001ew}).

Given the current and expected future progress in the NNLO
calculations~\cite{Misiak:2010dz}, reliable uncertainty estimates in the SM
prediction for ${\cal B}(\bar B\to X_s\gamma)$ can no longer be made without
evaluating the diagrams in Fig.~\ref{fig:tree} and checking what the actual
size of their contribution to the r.h.s of Eq.~(\ref{approx}) is. This fact
serves as the main motivation for our present work. Since the corrections are
expected (and found) to be quite small for $E_0=1.6\,$GeV, rough estimates of
their size are sufficient. Actually, nothing more is available within
perturbation theory alone because collinear logarithms $\ln(m_b/m_q)$
involving light quark masses $m_q$~ ($q=u,d,s$) remain in the final
expressions.  While using the so-called current masses for the light quarks is
not adequate in such a case, the above-mentioned rough estimates can be
obtained by assuming that $m_q$ are of the same order as masses of pions and
kaons or, equivalently, as the constituent quark masses. We shall do it by
varying $m_b/m_q$ from 10 to 50, which covers the necessary range. A refined
approach would require taking non-perturbative fragmentation into account, as
it has been done in Refs.~\cite{Kapustin:1995fk,Ferroglia:2010xe} for
contributions that are proportional to $|C_8|^2$ or in
Ref.~\cite{Ligeti:1999ea} for the $b \to u\bar u d\gamma$ background. Such
an analysis is beyond the scope of the present paper.

The article is organized as follows. Our final perturbative
results together with a discussion of their numerical impact on the
total decay rate are presented in Sec.~\ref{sec:results}. The
next two sections contain brief descriptions of two alternative methods that
we have used for integration over the four-body phase space. In
Sec.~\ref{sec:massive}, partly massive phase-space integration in $D=4$
dimensions is outlined. In Sec.~\ref{sec:dimreg}, a calculation involving
dimensional regularization and splitting functions is described. We
conclude in Sec.~\ref{sec:concl}.  App.~A contains a collection of
several intermediate results that may be useful for other calculations of
electromagnetic bremsstrahlung corrections to decays mediated by four-fermion
operators.  App.~B is devoted to summarizing the necessary properties of
the splitting functions.\\

\newsection{Results \label{sec:results}} 

\begin{figure}[b]
\begin{center}
\includegraphics[width=6cm,angle=0]{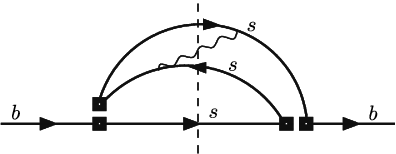}
\caption{\sf Sample contribution to the $b \to ss\bar s\gamma$ cross-terms
that give rise to the function $T_3(\delta)$. It is represented as a cut
propagator diagram, with the cut denoted by the vertical dashed line.
\label{fig:scrossed}}
\end{center}
\end{figure}

Our final result can be expressed in terms of three functions
$T_k(\delta)$~$(k=1,2,3$) that depend on the photon energy cut
$E_0=\f{m_b}{2}(1-\delta)$ and on logarithms of the quark masses. Each of the
functions gets multiplied by a quadratic polynomial in the Wilson coefficients
$C_i$ that are evaluated at the scale $\mu_b$. We assume that all the 
$C_i$ are real, as it is the case in the SM. The LO tree-level contribution to
$\Gamma[b\rightarrow X^p_s \gamma]$ arising from the four-quark operators
$P^u_{1,2}$ and $P_{3,\ldots, 6}\,$ reads
\begin{widetext}
\bea 
\Delta\Gamma[ b \to X_s^{p} \gamma]^{\mbox{\tiny LO}}_{\mbox{\tiny 4-quark}} &=& 
\f{G_F^2 \alpha_{\rm em}m_b^5}{32 \pi^4}\, \left| V^*_{ts} V_{tb} \right|^2 
\left[ T_1(\delta) \left(C_3^2 + 20 C_3 C_5 + \f{2}{9} C_4^2 
+ \f{40}{9}C_4 C_6 + 136 C_5^2 + \f{272}{9}C_6^2\right) \right.\nnb\\[1mm]
&&\hspace{-3cm}  
+\; T_2(\delta) \left(\f{2}{9} |A_1|^2 + |A_2|^2 
   + \left(\f{8}{9}C_3-\f{4}{27}C_4 + \f{128}{9} C_5 -\f{64}{27} C_6 \right) {\rm Re\,} A_1
+\left( \f{2}{3} C_3 + \f{8}{9} C_4 + \f{32}{3} C_5 + \f{128}{9} C_6 \right) {\rm Re\,} A_2 \right)\nnb\\[1mm]
&&\hspace{-3cm} \left. 
+\; T_3(\delta) \left( 
C_3^2 + \f83 C_3 C_4 + 32 C_3 C_5         + \f{128}{3} C_3 C_6 
      - \f29 C_4^2   + \f{128}{3} C_4 C_5 - \f{64}{9} C_4 C_6 
                     + 256 C_5^2          + \f{2048}{3} C_5 C_6 
                                          - \f{512}{9}C_6^2  \right) \right],\nnb\\ 
\label{pert.decay}
\eea
\ \\[-5mm]
where $A_i = -C_i \f{V^*_{us}V_{ub}}{V^*_{ts}V_{tb}},\, i=1,2$,~ and\\[-5mm]
\end{widetext}
\bea \label{tt1}
T_1(\delta) &=&\left(-\f{5}{3} \rho(\delta) - \f{2}{9}\omega(\delta)\right)
\ln\f{m_b^2\,\delta}{\sqrt[3]{m_u^4 m_d m_s}} \;+\; \f{109}{18}\delta\nnb\\[1mm]
&+& \f{17}{18}\delta^2 -\f{191}{108} \delta^3 + \f{23}{16} \delta^4 + \f{79}{18} \ln(1-\delta)\nnb\\[1mm] 
&-& \f{5}{3} \text{Li}_2(\delta) + \f{1}{9} \rho(\delta)\, \ln\f{m_s^5}{m_u^4 m_d}\,,
\eea
\bea \label{tt2}
T_2(\delta) &=& \left(-\f{1}{2} \rho(\delta) - \f{2}{27}\omega(\delta)\right)
\ln\f{m_b^2\,\delta}{m^2_{u}} \;+\; \f{187}{108}\delta\nnb\\[1mm]
&+& \f{7}{18}\delta^2 -\f{395}{648}\delta^3  + \f{1181}{2592}\delta^4
+ \f{133}{108} \ln(1-\delta)\nnb\\[1mm]
&-& \f{1}{2}\text{Li}_2(\delta) \,+\; \f{1}{9} \rho(\delta)\, \ln\f{m_s}{m_u}\,,
\eea 
\bea \label{tt3}
T_3(\delta) &=& \left(-\f{1}{18} \rho(\delta) - \f{1}{162}\omega(\delta)\right)
\ln\f{m_b^2\,\delta}{m^2_{s}} + \f{35}{162}\delta +\f{1}{72}\delta^2\nnb\\[1mm] 
&&\hspace{-1cm} -~ \f{89}{1944} \delta^3 +  \f{341}{7776} \delta^4 
+ \f{13}{81} \ln(1-\delta)-\f{1}{18} \text{Li}_2(\delta),~~~~~~~~
\eea
with
\bea
\rho(\delta) &=& \delta + \f{1}{6} \delta^4 + \ln(1-\delta),\nnb\\
\omega(\delta) &=& \f{3}{2} \delta^2 -2\delta^3 + \delta^4.
\eea
The function $T_3(\delta)$ originates from cross-terms in $b\to ss\bar
s\gamma$ where the $s$-quark lines are interchanged in one of the interfered
diagrams (see Fig.~\ref{fig:scrossed}). All the other contributions from
the penguin operators $P_{3,\ldots,6}$ alone are described by
$T_1(\delta)$. Finally, $T_2(\delta)$ comes from $P_{1,2}^u$ and their
interference with $P_{3,\ldots,6}$.

We have retained the light quark masses $m_q$ in the collinear
logarithms {\em only}, i.e.,  all the power-like corrections proportional to
$m_q^2/m_b^2$ have been neglected in the above expressions. Such an
approximation breaks down at some point, which manifests itself in
non-physical negative values of $T_{1,2}(\delta)$ when
$\ln(m_b^2\,\delta/m_q^2)$ is not big enough.
\begin{table}[t]
\begin{center}
\begin{tabular}{|r|r|r|r|}
\hline
$C_1 = -0.8144           $&$ C_3 = -0.0125 
$&$ C_5 = 0.0012 $&$ C_7 = -0.3688 $\\\hline
$C_2 = \phantom{-}1.0611 $&$ C_4 = -0.1224 
$&$ C_6 = 0.0026 $&$ C_8 = -0.1710$\\\hline
\end{tabular}
\caption{\sf The LO Wilson coefficients $C_i$~ 
at $\mu_b= 2.5\;{\rm GeV}$. The matching scale $\mu_0$
has been set to $160\,$GeV in their evaluation.\label{tab:Wilson}} 
\end{center}
\end{table}

Determining the size of the calculated correction is now straightforward.
Numerical values of the LO Wilson coefficients 
$C_i\,\equiv\,C_i^{(0)}(\mu_b)$ are summarized in
Tab.~\ref{tab:Wilson}. For the CKM element ratio we use
$\f{V^*_{us}V_{ub}}{V^*_{ts}V_{tb}} = -0.0079 +
0.018\,i$~\cite{Charles:2004jd}.  As far as the light quark masses $m_{u,d,s}$
are concerned, we set all of them equal in the numerical examples to be
discussed below.

In Tab.~\ref{tab:num}, we present the calculated correction~(\ref{pert.decay})
as a fraction of the leading contribution to the decay rate 
$\Gamma^{(0)} = G_F^2 \alpha_{\rm em}m_b^5\, C_7^2\, \left| V^*_{ts} V_{tb} \right|^2/(32 \pi^4)$
for two values of $E_0$ and two values of $\f{m_b}{m_q}$. Strong dependence on
the collinear logarithms is clearly visible. On the other hand, the
non-logarithmic terms turn out to be relevant in the considered range of
$\f{m_b}{m_q}$. For a low value of the photon energy cutoff $E_0 \simeq
m_b/20$ that has often been used in the literature, the correction can enhance
the inclusive rate by more than 10\%. On the other hand, for $E_0 = 1.6\,{\rm
GeV} \simeq \f{m_b}{3}$ or higher, the effect does not exceed $0.4\%$, which
is obviously due to phase-space suppression that becomes efficient when we
approach the high-energy endpoint $E_0 \simeq \f{m_b}{2}$, i.e. when
$\delta$ becomes small. In this limit, our correction in
Eq.~(\ref{pert.decay}) behaves like ${\cal O}\left(\delta^2\ln\delta\right)$. 
In all the four cases shown in Tab.~\ref{tab:num}, the
contribution from $T_1(\delta)$ is the dominant one, while $T_3(\delta)$
($T_2(\delta)$) give several times smaller effects. Thus, the
CKM-suppressed contributions that come with $T_2(\delta)$ are minuscule
indeed, and no precise knowledge of $\f{V^*_{us}V_{ub}}{V^*_{ts}V_{tb}}$ is
necessary here. On the other hand, our LO results exhibit significant
dependence on the renormalization scale $\mu_b$ that comes from
$C_{3,\ldots,6}(\mu_b)$. It could be stabilized only after including ${\cal
O}(\alpha_s)$ contributions to $b\to q\bar{q}s\gamma$, in particular the ones
generated by $P_{1,2}$~\cite{Huber:2014nna}.

\newsection{Calculation involving partly massive phase-space integrals\label{sec:massive}}

Let us now briefly describe the calculation. First, we consider
the diagrams in Fig.~\ref{fig:tree} with an operator 
\be \label{p3tilde}
\widetilde{P}_3 = (\bar{s}_L \gamma_{\mu} b_L)(\bar{q}_1\gamma^{\mu} q_2),
\ee
where no sum over flavors is present (contrary to $P_3$ in
Eq.~(\ref{physical})), and the electric charges $\{Q_s,Q_b,Q_1,Q_2\}$ are
retained arbitrary. The invariant matrix element ${\cal M}$ is calculated in
the Feynman gauge, so collinear divergences are allowed to occur in
interferences between different diagrams (rather than in self-interference terms
alone)~\cite{Gardi:2001di}.  The Dirac algebra is performed in $D=
4 -2\epsilon$ dimensions without neglecting the light quark masses.  At
this point, we proceed in two different ways.  One of them is to
integrate the spin-averaged $|\overline{\cal M}|^2$  over the
partly massive four-body phase space in $D=4$ dimensions.
The other way is to initially neglect the light quark masses, 
integrate $|\overline{\cal M}|^2$  in $D=4-2\epsilon$ dimensions over
the massless phase space, and convert the collinear $1/\epsilon$
poles to logarithms of masses only afterwards (see Sec.~\ref{sec:dimreg}).
\begin{table}[t]
\begin{center}
\begin{tabular}{|c|l|r|r|r|}
\hline
$ E_0            $&$ ~~~\delta $& \multicolumn{3}{c|}{
$\Delta\Gamma^{\mbox{\tiny LO}}_{\mbox{\tiny 4-quark}}/\Gamma^{(0)}$} \\\hline
$                $&$        $&$  $&$ \f{m_b}{m_q}=50 $&$ \f{m_b}{m_q}=10 $\\\hline
$ 1.6\,{\rm GeV} $&$ 0.316  $&$ -0.55\% +0.24\% \ln\f{m_b}{m_q} $&$ +0.40\% $&$ +0.01\% $\\\hline
$ \f{m_b}{20}    $&$ 0.9    $&$ -9.8\% + 6.1\% \ln\f{m_b}{m_q} $&$ +14.1\% $&$ +4.3\% $\\\hline
\end{tabular}
\caption{\sf The correction from Eq.~(\ref{pert.decay}) as a fraction of the
leading contribution $\Gamma^{(0)}$ for several values of $E_0$ and
$m_b/m_q$.\label{tab:num}}
\end{center}
\end{table}
  
In the $D=4$ case, we impose the photon energy cut (in the $b$-quark rest
frame ${\cal F}$) after expressing $E_\gamma$ in terms of invariants,
namely $2 m_b E_{\gamma} = m_b^2-s_{123}$, where $s_{123}$ is
the invariant mass squared of the $q_1\bar{q}_2 s$ system. A boost to the rest
frame $\hat{\cal F}$ of this system is performed along the $-\vec{k}$
direction, where $\vec{k}$ is the photon three-momentum in ${\cal F}$.  The
$q_1\bar{q}_2 s$ system has energy equal to $\sqrt{s_{123}}$ in $\hat{\cal
F}$, while the three-momenta of its constituents define a plane.  The
direction of $\vec{k}$ with respect to this plane is parametrized by
two polar angles $\hat\theta$ and $\hat\varphi$. The remaining
three phase-space variables are the quark energy fractions $\hat{x}_i =
\f{2\hat{E}_i}{\sqrt{s_{123}}}$ ($i=1,2$) and $s_{123}$.

We integrate first over $\hat\theta$ and 
$\hat\varphi$, and obtain results containing large logarithms
$\ln \f{m_b^2}{m_q^2}$.  Integration of the interference terms
$\f{p_i p_j}{(p_i q) (p_j k)}$ involves angular ordering (see
Ref.~\cite{Ellis:2003Cam}), which applies also to the case
$p_i^{2}\equiv m_q^2\ll m_b^2$.  At this 
stage, all the collinear logarithms have been already
identified, which allows us to neglect masses of the outgoing
particles in the remaining terms. Next, integrations over 
the quark energy fractions and $s_{123}$ are performed.
Finally, charge conservation $Q_b = Q_1 - Q_2+Q_s$ is
imposed, but the charges on the r.h.s. are still retained
arbitrary. Several intermediate results with such arbitrary charges are
collected in App.~A.

Once the final result for $\widetilde{P}_3$ with arbitrary charges is at hand,
obtaining the corresponding one for $P_3$ ($T_1(\delta)$) is just a matter of
substituting the actual values of the charges and summing over flavors.
Extending the calculation to $P_{4,5,6}$ and taking into account the $b\to
ss\bar s\gamma$ cross-terms ($T_3(\delta)$) requires modifying the Dirac
algebra and color factors but no essential difference in the phase-space
integration is encountered. As far as $T_2(\delta)$ is concerned, it
originates from the operators $P_{1,2}^u$, including their interference with
$P_{3,\ldots,6}$. In this case, it is sufficient to express $P_{1,2}^u$ in $D=4$
dimensions as linear combinations of the $(\bar s b)(\bar u u)$-parts of
$P_{3,\ldots,6}$, namely
\bea 
P^u_1 ~= & -\f{4}{27} P^u_3 +\f{1}{9} P^u_4 +\f{1}{27} P^u_5 -\f{1}{36} P^u_6 \nnb\\[1mm]
P^u_2 ~= & -\f{1}{9} P^u_3 -\f{2}{3} P^u_4 +\f{1}{36} P^u_5 +\f{1}{6} P^u_6 \;,
\label{Fierz1u}
\eea
The above expressions are easily derived using Fierz identities for 
the Dirac and Gell-Mann matrices.

\newsection{Calculation with the help of dimensional regularization\label{sec:dimreg}}

A calculation with the help of dimensional regularization is technically
simpler but involves a few subtleties. To start with, all the particles in the
final state are assumed to be massless, and the phase-space integration is
performed in $D=4-2\epsilon$ dimensions~\cite{Kaminski:2007MSc}. The
results are collected in App.~A. Given that the collinear divergences
appear as $1/\epsilon$ poles, one should make sure that no 
ambiguities arise from Dirac traces with odd numbers of
$\gamma_5$'s~\cite{Chetyrkin:1997gb}. Fortunately, such traces being purely
imaginary give no contribution to the decay rate in the case of $P_{3,\ldots,6}$
alone. As far as $P_{1,2}^u$ are concerned, we do not need to consider them at
this level.  Once the collinear divergences are re-expressed in terms of
logarithms of masses (see below), we can pass to $D=4$ and use the identities
(\ref{Fierz1u}).

For definiteness, let us consider the operator $\widetilde{P}_3$ from
Eq.~(\ref{p3tilde}) again, but this time with the $s$-quark denoted by $q_3$
to keep the notation symmetric. Before integrating over the photon energy (but
after integration over all the other phase-space variables), the differential
decay width for $b \to q_1\bar{q}_2 q_3 \gamma$ reads
\be \label{diffrate}
\f{d\Gamma}{dx} = \f{d\Gamma_{\epsilon}}{dx} + \f{d\Gamma_{\rm shift}}{dx}\,,
\ee
where $x=2E_\gamma/m_b$. The first term on the r.h.s. above is the
dimensionally regulated expression, while the second one converts the
dimensional regulators to logarithms of masses. Its explicit form is given
below.

We shall need a $D$-dimensional expression for the total width of the three-body
decay $b \to q_1\bar{q}_2 q_3$. Denoting momenta of the final-state
massless particles by $p_i$, $i=1,2,3$, and parametrizing the phase space by
$s_{ij} = 2p_i p_j/m_b^2$, one can write
\be
\Gamma_{\rm 3\mbox{-}body} = \f{\widetilde{\mu}^{4\epsilon}}{2 m_b}
\int_0^1 \!ds_{12}\,ds_{13}\,ds_{23}\,|\overline{\cal M}{}'|^2\, M_3,  
\ee
where ${\cal M}{}'$ is the corresponding invariant matrix element,
$\widetilde{\mu}^2 = \mu^2 e^{\gamma_E}/4\pi$, and
\be \label{eq:m3}
M_3 = \f{ \delta (1-s_{12}-s_{13}-s_{23})\;
\pi^{-\f{5}{2} + 2\epsilon}\, (m_b^2)^{1-2\epsilon}}{ 2^{8-6\epsilon}\, 
\Gamma\left(\f{3}{2}-\epsilon\right)\, \Gamma(1-\epsilon)\, (s_{12}s_{13}s_{23})^\epsilon}
\ee
describes the phase-space measure~\cite{GehrmannDeRidder:2003bm}. 
The second term in Eq.~(\ref{diffrate}) can now be written as
\bea
\f{d\Gamma_{\rm shift}}{dx} &=& \f{\widetilde{\mu}^{4\epsilon}}{2 m_b}
\int_0^1 \!ds_{12}\,ds_{13}\,ds_{23}\,|\overline{\cal M}{}'|^2\, M_3\; \f{\alpha_{\rm em}}{2\pi x}\nnb\\
&\times& 
\left\{ Q_1^2 \left[ 1 + \left(1-\f{x}{1-s_{23}}\right)^2\right] \Theta\left(1-s_{23}-x\right)\right.\nnb\\
&\times& \left. \left[ \f{1}{\epsilon} - 1 + 2\ln\f{(1-s_{23})\mu}{xm_{q_1}}\right]
+\mbox{(cyclic)}\right\}. \label{shift}
\eea
Properties of the splitting functions that have been necessary to derive the
above formula are summarized in App.~B. One should remember that all the
collinear $1/\epsilon$ poles cancel in Eq.~(\ref{diffrate}) only after the 
charge conservation $Q_b = Q_1-Q_2+Q_s$ has been imposed.

The structure of Eq.~(\ref{shift}) remains the same irrespective of what
interaction generates the $b$-quark decay.  Thus, it is applicable as it
stands to the operators $P_{3,\ldots,6}$. Next, as already mentioned,
Eq.~(\ref{Fierz1u}) is used in $D=4$ dimensions to take $P_{1,2}^u$
into account. This way the final expression in Eq.~(\ref{pert.decay}) has been
obtained once again, in perfect agreement with the results of
Sec.~\ref{sec:massive}.

\newsection{Conclusions \label{sec:concl}}

In the present paper, we have evaluated the LO contributions to the partonic
decay width $\Gamma(b\to X_s^p\gamma)$ that originate from the four-quark
operators $P_{1,2}^u$ and $P_{3,\ldots,6}$. They can be sizeable (above 10\%) for
low photon energy cutoffs $E_0$ but become very small (below 0.4\%) in the
phenomenologically interesting domain $E_0 \geq 1.6\,$GeV, i.e., for $\delta
\equiv 1-2E_0/m_b \lsim 0.32$.  For small $\delta$, they behave like ${\cal
O}(\delta^2\ln\delta)$, which determines their phase-space suppression near the
endpoint.

The presence of collinear logarithms $\ln(m_b^2 \delta/m_q^2)$ involving light
quark masses $m_q$ ($q=u,d,s$) implies that our perturbative results (with
$m_b/m_q$ varied from 10 to 50) may serve only as rough estimates of the
corresponding contributions to the inclusive hadronic $\bar B \to X_s\gamma$
decay rate. Having such rough estimates at hand is both advantageous and
sufficient at present, given that the overall non-perturbative uncertainty
remains at the $\pm 5\%$ level~\cite{Benzke:2010js}. However, once our
control over non-perturbative corrections improves in the future, the current
contributions will need to be supplemented with hadronic fragmentation effects,
along the lines of Refs.~\cite{Kapustin:1995fk,Ferroglia:2010xe}.

\newappendix{ACKNOWLEDGMENTS}

We thank Einan Gardi, Thomas Gehrmann and Tobias Huber for helpful
discussions. We are grateful to Tobias Huber and Javier Virto for pointing out
missing factors of 2 in the terms proportional to $T_3(\delta)$ in an earlier
version of this paper, as well as for confirming our updated
results~\cite{Huber:2014nna}. This work has been supported in part by the
National Science Centre (Poland) research project, decision no
DEC-2011/01/B/ST2/00438.  M.P. acknowledges partial support from the DFG under
project MA 1187/10-2 ``Theoretische Untersuchung von effektiven Feldtheorien
f\"ur Pr\"azisionsvorhersagen von $B$-Zerf\"allen''.  M.M. acknowledges
support from the DFG through the ``Mercator'' guest professorship program.

\newappendix{APPENDIX A: INTERMEDIATE RESULTS \label{app:interm}}
\def\theequation{A.\arabic{equation}}

Here, we present several intermediate results that might
be useful for studying hard photon emission in other processes
mediated by four-fermion operators, like muon decays or
semileptonic heavy quark decays. Although radiative corrections to such
processes have been calculated long ago, none of the published results that we are
aware of leaves the photon energy as the only phase-space variable that has not been
integrated over (or, equivalently, integrated with an arbitrary cutoff).
To make our results applicable to such cases, it is enough to present them for
arbitrary electric charges of the final-state fermions.

As in Sec.~\ref{sec:massive}, we replace $P_3$ (\ref{physical}) by
$\widetilde{P}_3$ (\ref{p3tilde}) and assume that $q_1 \neq s$, which means
that no cross-terms ($T_3(\delta)$) arise. The contribution of
$\widetilde{P}_3$ to the decay rate is as in the $C_3^2\, T_1(\delta)$ term in
Eq.~(\ref{pert.decay}) but with $T_1(\delta)$ replaced by
\bea 
\widetilde{T}_1(\delta)	&=& 
(Q_1^2 + Q_2^2) \, F_{11}(\delta) + Q_{s}^2\,  F_{ss}(\delta)\nnb\\[1mm] 
&+& Q_1 Q_2 \left( -\f{5}{6}\rho(\delta) -\f{1}{9} \omega(\delta)-\f{1}{4}\delta^2 \right)\nnb\\[1mm]
&+& (Q_1 - Q_2 )\,Q_s \left( \f{7}{6}\rho(\delta) +\f{1}{18} \omega(\delta) \right)
\,,~~~~~ \label{eq:T1tilde}
\eea
where 
\bea 
F_{11}(\delta) &=& \left(-\rho(\delta) - \f{1}{6}\omega(\delta)\right)
\ln\f{m_b^2\,\delta}{m_q^2} + 4\delta +\f{11}{12}\delta^2\nnb\\ &-&\f{17}{12} \delta^3 
+ \f{79}{72} \delta^4 
+ 3 \ln(1-\delta)- \text{Li}_2(\delta)\,,\nnb\\[2mm]
F_{ss}(\delta) &=& -\, \rho(\delta)\, \ln\f{m_b^2\,\delta}{m_q^2}\;
+\f{23}{6}\delta -\f{1}{12}\delta^3 + \f{61}{144}\delta^4\nnb\\
&+& \f{17}{6} \ln(1-\delta) -\text{Li}_2(\delta)\,.
\label{eq:partmassresults1}
\eea 
In Eq.~(\ref{eq:T1tilde}), charge conservation $Q_b=Q_1-Q_2 + Q_s$ has been
already imposed. In effect, collinear logarithms remain only in the terms that
come with $Q_i^2,\,i=1,2,s$.  For simplicity, all the light quark masses have
been set equal and denoted by $m_q$. However, it is easy to relax this
assumption and identify them by the corresponding charges despite the fact
that charge conservation has already been used (see App.~B).

Now, let us consider the case when $q_1 = s$ in Eq.~(\ref{p3tilde}). Then,
apart from $\widetilde{T}_1(\delta)$~(\ref{eq:T1tilde}), we get an additional contribution
from the cross-terms
\bea
\widetilde{T}_3(\delta) &=& \f12 Q_{2}^2\, S_{22}(\delta) + \f12 Q_{s}^2\,S_{ss}(\delta)\nnb\\ 
&+& Q_{2} Q_{s} \left( -\f{5}{18}\rho(\delta) -\f{1}{27} \omega(\delta)-\f{1}{12}\delta^2 \right)
\label{eq:T3tilde}
\eea
where
\bea 
S_{22}(\delta) &=&\left( -\f{1}{3} \rho(\delta) - \f{1}{9}\omega(\delta)\right)
\ln\f{m_b^2\,\delta}{m_q^2} + \f{25}{18}\delta +\f{11}{18}\delta^2\nnb\\ 
&-&\f{11}{12} \delta^3 + \f{85}{144} \delta^4 + \f{19}{18} \ln(1-\delta)
- \f{1}{3}\text{Li}_2(\delta)\,,\nnb\\[2mm]
S_{ss}(\delta) &=& -\f{2}{3}\,\rho(\delta)\,\ln\f{m_b^2\,\delta}{m_q^2}\;
+\f{55}{18}\delta -\f{1}{12}\delta^2-\f{1}{18}\delta^3\nnb\\ 
&+& \f{79}{216}\delta^4 +\f{43}{18} \ln(1-\delta)
-\f{2}{3}\text{Li}_2(\delta)\,.\label{eq:partmassresults2}
\eea

The above results are valid only for the particular Dirac structure of the
operator $\widetilde{P}_3$. To generalize them to all the four-fermion
operators with chirality-conserving currents, it is sufficient to consider 
\be
\widetilde{P}_5 = (\bar{s}_L \gamma_{\mu_1} \gamma_{\mu_2} \gamma_{\mu_3} b_L) 
                         (\bar{q}_1 \gamma^{\mu_1} \gamma^{\mu_2}\gamma^{\mu_3} q_2). 
\ee
Its interference with $\widetilde{P}_3$ gives 
\be \label{xtra35}
20\, \widetilde{T}_1(\delta) + (Q_1^2-Q_2^2)\,r_1(\delta) + (Q_1+Q_2)\,Q_s\,r_2(\delta), 
\ee
and $\,32\,\widetilde{T}_3(\delta)\,$ for the cross-terms. Its 
self-interference gives 
\bea
136\,\widetilde{T}_1(\delta) + 10\left[(Q_1^2-Q_2^2)\,r_1(\delta) + (Q_1+Q_2)\,Q_s\,r_2(\delta)\right],\nnb\\
\label{xtra55}
\eea
and $\,256\,\widetilde{T}_3(\delta)\,$ for the cross-terms. The functions $r_{1,2}(\delta)$ are given by
\bea 
r_1(\delta) &=& 2\omega(\delta) \ln\f{m_b^2\,\delta}{m_q^2}
-2\rho(\delta)-11\delta^2 + 16 \delta^3 - \f{31}{4} \delta^4\,, \nnb\\[2mm]
r_2(\delta) &=& 4\rho(\delta)-\f{2}{3}\omega(\delta)-3\delta^2\,. \label{eq:restfuns}
\eea 

The corresponding results for massless final-state fermions in the dimensional
regularization have been obtained with the help of Eq.~(A.5) of
Ref.~\cite{GehrmannDeRidder:2003bm} where a compact expression for the
$D$-dimensional four-body phase-space measure is given. Only the functions that
come proportional to squared charges differ from their $D=4$
counterparts. For $\widetilde{P}_3$ alone, they read
\bea 
F_{11}^{(\epsilon)}(\delta) &=&
\left( \rho(\delta) + \f{1}{6}\omega(\delta)\right) C_\epsilon(\delta)
+ \f{45}{4}\delta +\f{5}{2}\delta^2 -\f{23}{9} \delta^3\nnb\\ 
&+& \f{175}{72} \delta^4 +\f{37}{4}\ln(1-\delta)
+ \ln^2(1-\delta)- 2\text{Li}_2(\delta)\,,\nnb\\[2mm]
F_{ss}^{(\epsilon)}(\delta) &=& \rho(\delta)\, C_\epsilon(\delta)
+\f{133}{12}\delta +\f{7}{24}\delta^2 + \f{4}{9}\delta^3 + \f{65}{72}\delta^4\nnb\\
&+&\f{109}{12} \ln(1-\delta)
+\ln^2(1-\delta) -2\text{Li}_2(\delta)\,,\nnb\\[2mm]
S_{22}^{(\epsilon)}(\delta) &=&
\left( \f{1}{3} \rho(\delta) + \f{1}{9}\omega(\delta)\right) C_\epsilon(\delta)
+ \f{125}{36}\delta +\f{101}{72}\delta^2\nnb\\ 
&-&\f{44}{27} \delta^3 + \f{83}{72} \delta^4 +\f{101}{36} \ln(1-\delta) 
+ \f{1}{3}\ln^2(1-\delta)\nnb\\
&-& \f{2}{3}\text{Li}_2(\delta)\,,\nnb\\[2mm]
S_{ss}^{(\epsilon)}(\delta) &=& \f{2}{3}\,\rho(\delta)\, C_\epsilon(\delta)
+\f{65}{9}\delta +\f{1}{9}\delta^2 + \f{8}{27}\delta^3 + \f{31}{54}\delta^4\nnb\\
&+&\f{53}{9} \ln(1-\delta) +\f{2}{3} \ln^2(1-\delta) -\f{4}{3}\text{Li}_2(\delta)\,,
\label{fsdimreg}
\eea
where $C_\epsilon(\delta) = 1/\epsilon - 2\ln[\delta(1-\delta)]$.

In the case of $\widetilde{P}_3$-$\widetilde{P}_5$ interference
(Eq.~(\ref{xtra35}) and below), the following replacements need to be made:
\bea 
(Q_1^2-Q_2^2)\,r_1 &\to& 
-\,2 (Q_1^2-Q_2^2) \left( (C_\epsilon+8)\,\omega +7 \,\rho - \f{\delta^2}{4} \right) \nnb\\
&& -\,12\,Q_s^2\,\rho -4\, Q_2^2 (6\rho + \omega)\,,\nnb\\[2mm] 
32\,\widetilde{T}_3 &\to& 32\,\widetilde{T}_3^{(\epsilon)} 
-\f{8}{9} Q_s^2 ( 3\rho -\omega) -\f{4}{3} Q_2^2 ( \rho +\omega)\,,\nnb\\
\label{repl35}
\eea
where $\widetilde{T}_3^{(\epsilon)}(\delta)$ is given in terms of
$S_{22}^{(\epsilon)}(\delta)$ and $S_{ss}^{(\epsilon)}(\delta)$, in analogy to
Eq.~(\ref{eq:T3tilde}).  
The corresponding replacements for the $\widetilde{P}_5$ self-interference
(Eq.~(\ref{xtra55}) and below) read
\bea 
&& \hspace{-4mm} (Q_1^2-Q_2^2)\,r_1(\delta) \;\to \nnb\\
&& \hspace{15mm} -\,2 (Q_1^2-Q_2^2) \left( \left( C_\epsilon + \f{83}{10}\right)\omega
+10 \,\rho - \f{\delta^2}{4}\right)\nnb\\
&& \hspace{15mm} +\, Q_s^2 \left(- 18\rho +\f{8}{5}\omega\right)
-4\, Q_2^2 \left( 9 \rho + \f{19}{10} \omega \right)\,,\nnb\\[2mm]
&& \hspace{-4mm} 256\,\widetilde{T}_3 \;\to\; 256\,\widetilde{T}_3^{(\epsilon)} 
-\f{32}{9} Q_s^2 ( 21\rho -4\,\omega) -\f{16}{3} Q_2^2 ( 7\rho +5\omega)\,.\nnb\\
\label{repl55}
\eea
Conversion of the collinear regulators in Eq.~(\ref{diffrate}) is most
conveniently performed before substitution of charges, at the level of 
the functions from Eqs.~(\ref{fsdimreg})--(\ref{repl55}).

Our final result in Eq.~(\ref{pert.decay}) has been obtained by forming
appropriate linear combinations of the functions appearing in
Eqs.~(\ref{eq:T1tilde}), (\ref{eq:T3tilde}) and (\ref{eq:restfuns}),
according to the values of charges, color factors and sums over flavors.

\newappendix{APPENDIX B: SPLITTING FUNCTIONS}
\def\theequation{B.\arabic{equation}}

Our conversion formula (\ref{shift}) involves a difference between
splitting functions derived in the dimensional regularization and
in the regularization with masses.  Their derivation along the lines
of Refs.~\cite{Cacciari:2001cw,Cacciari:2002xb} is briefly
described in the following. 

Let us consider an amplitude ${\cal M}(q,k;\ldots)$ of a process
where an external massive fermion radiates a photon with momentum
$k$~($k^2=0$). After the radiation, the fermion is on shell
($q^2=m^2$). We shall assume that the calculation is performed in the
light-cone axial gauge $n\cdot A = 0$, where $n$ is a lightlike vector ($n^2=0$),
and the sum over photon polarizations gives
\be
\sum\limits_{\lambda}\epsilon_{\mu}^{\lambda *}\epsilon_{\nu}^{\lambda} = 
-g_{\mu\nu} + \f{k_{\mu}n_{\nu} + k_{\nu}n_{\mu}}{kn}\,.
\ee
In such a gauge, interference terms between different diagrams are free 
of collinear singularities~\cite{Gardi:2001di}.

\begin{figure}[t]
\begin{center}
\includegraphics[width=8cm,angle=0]{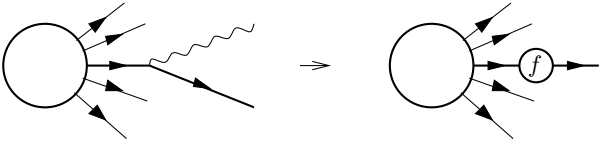}\\
\begin{picture}(0,0)
\put(86,37){$f$}
\end{picture}
\caption{\sf A schematic picture of the factorization formula (\ref{M2_dominant}) \label{fig:splitting}}
\end{center}
\end{figure}

In the course of the splitting function derivation, it is convenient to
introduce the Sudakov parametrization in terms of
\bea
p &\equiv& q \,+\, k \,-\, \f{qk}{(q+k)n}\; n,\nnb\\[2mm]
k_\perp &\equiv& z\,p \,-\, q \,+\, z(1-z)\, \f{(q+k)q}{qn}\; n,
\label{Sudakov_def}
\eea
where $z\equiv (qn)/[(q+k)n] \in \left[m^2/(m^2 +
2qk)\,,1\right]$.  It is easy to verify that $\,p^2 = m^2$, and that the
spacelike vector $k_{\perp}$ is orthogonal to both $n$ and $p$.  
Inverting the relations (\ref{Sudakov_def}), one finds
\bea
\label{Sudakov_par}
q &=&    z \,p \,-\, k_{\perp} \,+\, \f{\bold{k}_{\perp}^2 + (1-z^2) m^2}{z\,(2pn)}\;n,\nnb\\[2mm]
k &=& (1-z)\,p \,+\, k_{\perp} \,+\, \f{\bold{k}_{\perp}^2 - (1-z)^2 m^2}{(1-z)\,(2pn)}\;n,
\eea
where $\bold{k}_{\perp}^2$ stands for $-k_{\perp}^{\mu}k_{\perp\mu}$
when expressed in a frame-independent way.  In such a
parametrization, propagator denominators that are responsible for
collinear singularities appear as
\be
\label{prop}
\f{1}{(q+k)^2-m^2}=\f{1}{2qk}=
\f{z(1-z)}{\bold{k}_{\perp}^2+(1-z)^2 m^2}.
\ee
The transverse momentum $k_{\perp}$ parametrizes how far off-shell the
radiating fermion is.  In the massless case ($m=0$ and $\epsilon\neq 0$), the
collinear limit is defined by $k_{\perp}\rightarrow 0$, which 
determines the phase-space region where the $1/\epsilon$ singularity
arises.  In the case of a massive fermion ($m\neq 0$ and $\epsilon=0$), the
quasi-collinear limit has to be considered~\cite{Catani:2000ef}. In this
limit, the collinear region is defined by taking simultaneously
$\bold{k}_{\perp}^2\rightarrow 0$ and $m\rightarrow 0$, but keeping the ratio
$m^2/\bold{k}_{\perp}^2$ fixed. Both limits lead to the factorization
formula~\cite{Catani:1996vz,Cacciari:2001cw} illustrated in
Fig.~\ref{fig:splitting}
\be\label{M2_dominant}
|\overline{\cal M}(q,k;\ldots)|^2
\;\simeq\; \f{Q^2_j}{2qk}\,\hat{P}(z)\,
|\overline{\cal M}(p;\ldots)|^2\,,
\ee
where ${\cal M}(p;\ldots)$ is the amplitude of the process without
radiation where $p\rightarrow q/z$ in the collinear or quasi-collinear
limits, and $Q_j$ is the fermion charge. The splitting function 
$\hat{P}(z)$ in the collinear limit and in $D=4-2\epsilon$ dimensions reads
(see Eqs.~(53) and (54) of Ref.~\cite{Cacciari:2001cw})
\be
\hat{P}_{\epsilon}(z)=8\pi\, \alpha_e\left[\f{1+z^2}{1-z\;} -\epsilon\,(1-z)\right].
\ee
It becomes the Altarelli-Parisi splitting function~\cite{Altarelli:1977zs}
for the gluon emission off quark when $8\pi\alpha_e$ is
replaced by $C_F$. Its extension to the massive quark case in $D=4$ dimensions is 
\be
\hat{P}_{m}(z)= 8\pi\, \alpha_e\left[\f{1+z^2}{1-z\;}  -\f{m^2}{qk}\right]\,.
\ee

For definiteness, let us assume that the number of final-state particles is
as in Fig~\ref{fig:tree}. In the collinear region, it is possible to
disentangle the four-body phase space $d\Phi_4$ into a
convolution of the three-body phase space of the non-radiative
process, and the phase space corresponding to the radiation process
alone~\cite{Catani:1996vz,Dittmaier:1999mb},
\be
d\Phi_4 = d\Phi_3 \otimes d\Phi\,.
\ee
One proceeds with integration over the low-$k_{\perp}$ region using the
following phase-space measure~\cite{Cacciari:2001cw}
\bea
d\Phi &=& (\tilde{\mu})^{2\epsilon} \f{d^d k}{(2\pi)^{d-1}} \delta_{+}(k^2)=\nnb\\
&=&  \f{1}{16\pi^2}\f{1}{\Gamma(1-\epsilon)}
d\bold{k}_{\perp}^2 \left(\f{4\pi\tilde{\mu}^2}{\bold{k}_{\perp}^2} \right)^{\epsilon} \f{dz}{z(1-z)} \Theta(z(1-z))\,.\nnb\\
\eea
The integration is performed from $\bold{k}_{\perp}^2 = 0$ up to $\bold{k}_{\perp}^2
= E^2$, where $E$ is chosen to remain in the low-$k_{\perp}$ region, to
preserve the factorization formula.

The splitting functions integrated over $\bold{k}_{\perp}^2$ read
\bea
f_{\epsilon}(z) &=&\f{\alpha_e}{\pi} \left[\f{1+z^2}{1-z\;}\left( -\f{1}{2\epsilon} 
+ \ln\f{E}{\mu}  \right) +\f{1-z}{2} \right]\,,\nnb\\[4mm]
f_{m}(z) &=&  \f{\alpha_e}{\pi} \left[\f{1+z^2}{1-z\;} \ln\f{E}{(1-z)m}  - \f{z}{1-z}\right]\,,
\label{eqs:splitfunctions}
\eea 
while their difference is
\bea\label{splitfundiff}
\Delta f(z) &\equiv&f_{m}(z) -f_{\epsilon}(z) \nnb\\
&=&\f{\alpha_e}{2\pi}\, \f{1+z^2}{1-z\;} \left[ \f{1}{\epsilon} -1 -2\ln\f{(1-z) m}{\mu} \right].~~~~~~~~\\
\ \nnb
\eea
The dependence on $E$ cancels in Eq.~(\ref{splitfundiff}) because both
splitting functions in Eq.~(\ref{eqs:splitfunctions}) have been
consistently derived in the corresponding regularizations, and
they contain the same high-$k_{\perp}$ finite terms. The formula
(\ref{splitfundiff}) could have also been obtained with the splitting
functions from Ref.~\cite{Huber:2005ig}.

The mass-regulated and dimensionally regulated radiative decay widths
with $E_\gamma > E_0$ satisfy the following relation
\bea
\Gamma_m &=& \Gamma_\epsilon +
\sum_j\!Q^2_j\!\int d\widetilde{\Phi}_3 \int_0^1 dz\,\Delta f(z)\, 
|\overline{\cal M}(p;\ldots)|^2\nnb\\[1mm]
&\times& \Theta\left[(1-z)\, p^0_j - E_0\right]\,,
\label{eq:split.translation}
\eea
where the sum goes over all the radiating fermions.  The
three-body final-state phase-space measure $d\widetilde{\Phi}_3$
integrated over the ``blind'' angular variables 
reads~\cite{GehrmannDeRidder:2003bm}
\be
d\widetilde{\Phi}_3 \equiv \tilde{\mu}^{4\epsilon} \int_{\Omega} d\Phi_3 = 
\tilde{\mu}^{4\epsilon}\, M_3\, ds_{12}\, ds_{13}\, ds_{23}\,,
\ee
where $M_3$ has been given in Eq.~(\ref{eq:m3}).

Eq.~(\ref{eq:split.translation}) has been derived in the light-cone axial
gauge but it is actually gauge-independent once charge conservation has been
imposed. Thanks to this fact, we could have used it in our Feynman-gauge
calculation. Actually, the conversion formula (\ref{shift}) is obtained from
Eq.~(\ref{eq:split.translation}) just by differentiation with respect to 
$E_0$. Even after imposing charge conservation, the splitting functions
derived in the same regularization may differ by finite terms which depend on
the chosen gauge and also on the high-$k_{\perp}$ integration limit.
Only the difference (\ref{splitfundiff}) of the two splitting functions
is gauge- and convention-independent.

\end{document}